\documentclass[conference,10pt]{IEEEtran}
\IEEEoverridecommandlockouts
\usepackage{cite}

\usepackage{amsmath,amssymb,amsfonts, amsthm}

\usepackage{algorithm}
\usepackage{algorithmic}
\usepackage{graphicx}
\usepackage{textcomp}
\usepackage{xcolor}
\usepackage{bm}
\usepackage{bbm}
\usepackage{subcaption}
\def \bf{\mathbf}

\def\BibTeX{{\rm B\kern-.05em{\sc i\kern-.025em b}\kern-.08em
    T\kern-.1667em\lower.7ex\hbox{E}\kern-.125emX}}

\newtheorem{definition}{Definition}
\newtheorem{theorem}{Theorem}
\newtheorem{lemma}{Lemma}
\newtheorem{prop}{Proposition}

\begin{document}

\title{Age-Based Scheduling for a Memory-Constrained Quantum Switch}

\author{Stavros Mitrolaris \qquad Subhankar Banerjee \qquad Sennur Ulukus\\
\normalsize Department of Electrical and Computer Engineering\\
\normalsize University of Maryland, College Park, MD 20742\\
\normalsize \emph{stavros@umd.edu} \qquad \emph{sbanerje@umd.edu} \qquad \emph{ulukus@umd.edu}}

\maketitle

\begin{abstract}
In a time-slotted system, we study the problem of scheduling multipartite entanglement requests in a quantum switch with a finite number of quantum memory registers. Specifically, we consider probabilistic link-level entanglement (LLE) generation for each user,  probabilistic entanglement swapping, and one-slot decoherence. To evaluate the performance of the proposed scheduling policies, we introduce a novel age-based metric, coined \emph{age of entanglement establishment (AoEE)}. We consider two families of low-complexity policies for which we obtain closed-form expressions for their corresponding AoEE performance. Optimizing over each family, we obtain two policies. Further, we propose one more low-complexity policy and provide its performance guarantee. Finally, we numerically compare the performance of the proposed policies.
\end{abstract}

\section{Introduction}
In emerging applications of practical importance such as quantum key distribution \cite{ekert1991quantum}, which enables information-theoretically secure communication, and distributed quantum computing \cite{barral2025review, main2025distributed}, remote parties must share entangled qubits \cite{valls2024brief}. To meet this need, quantum switches have been proposed as central network nodes that coordinate the efficient establishment of entanglement among users. Specifically, the switch can establish end-to-end entanglement between users by first generating entanglement with users individually.   

\begin{figure}[t]
    \centering
    \begin{subfigure}[t]{0.27\linewidth}
        \centering
        \includegraphics[width=\linewidth]{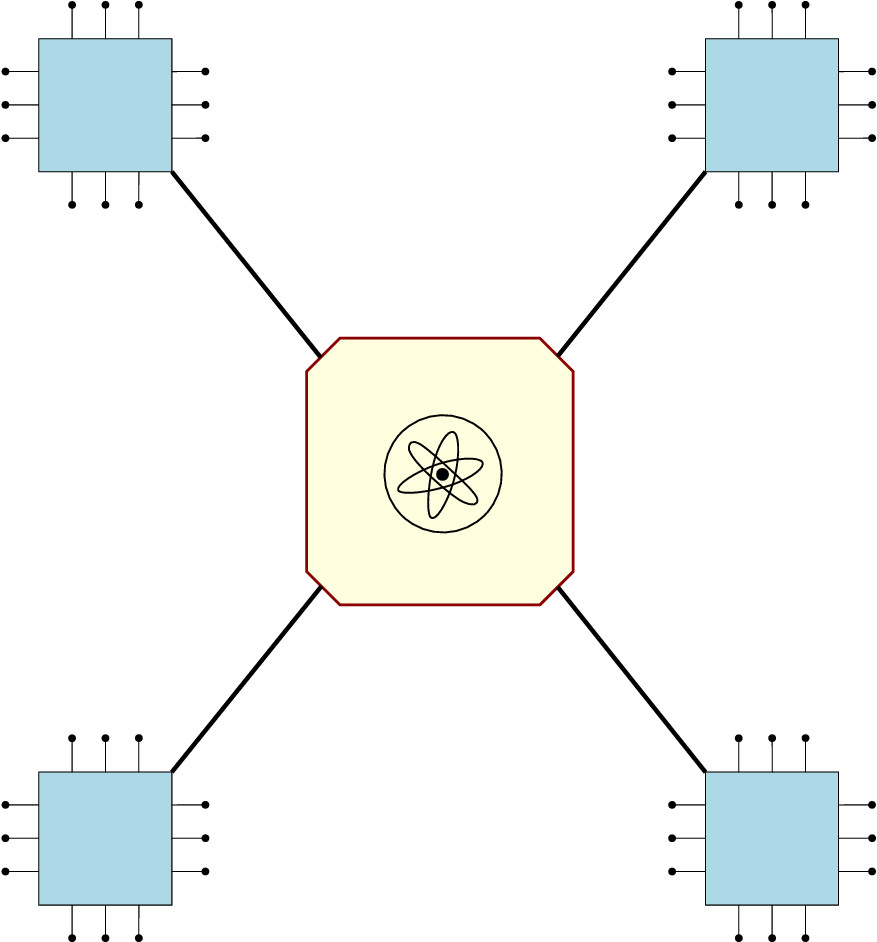}
        \caption{}
    \end{subfigure}\hfill
    \begin{subfigure}[t]{0.27\linewidth}
        \centering
        \includegraphics[width=\linewidth]{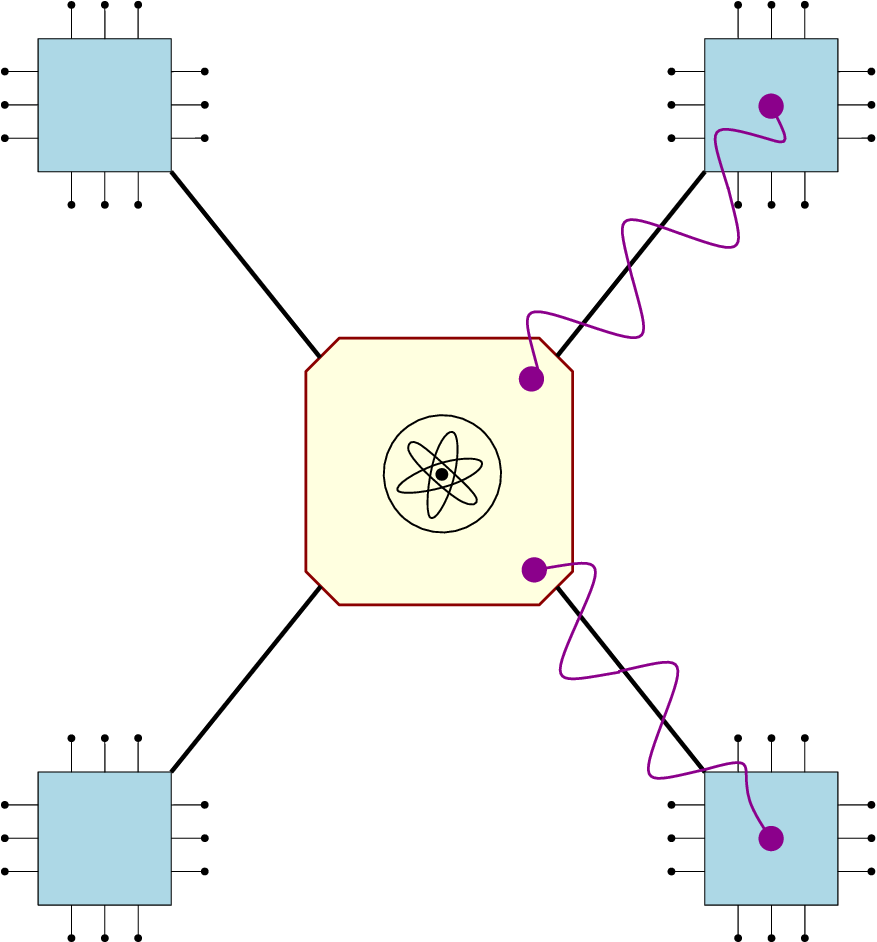}
        \caption{}
    \end{subfigure}\hfill
    \begin{subfigure}[t]{0.27\linewidth}
        \centering
        \includegraphics[width=\linewidth]{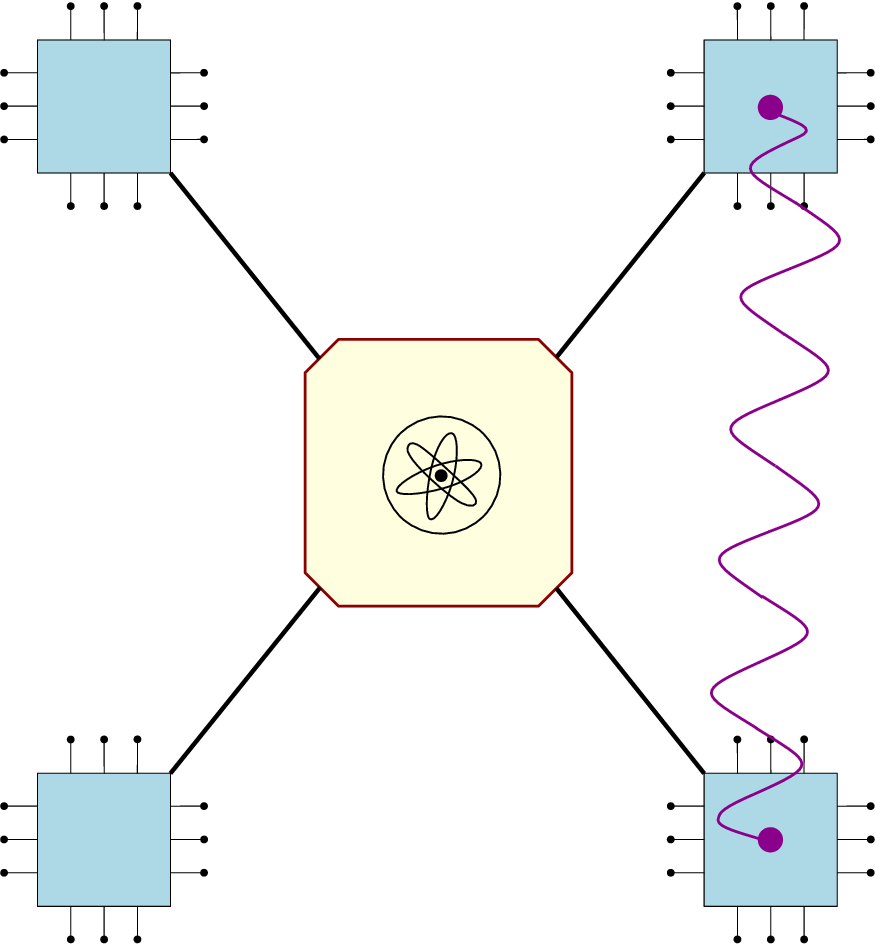}
        \caption{}
    \end{subfigure}
    \caption{An illustrative example of an entanglement swapping operation. In (a), the quantum switch is connected to four users, and no LLEs have been established. In (b), the switch establishes LLEs with two users and stores the corresponding qubits in the memory. In (c), by performing local operations on the stored qubits, the switch creates an end-to-end entangled pair between the two users.}
    \label{fig:swapping}
\end{figure}

In particular, a quantum switch is connected to each user via a dedicated link that supports the generation of link-level entanglements (LLEs). An LLE corresponds to a pair of entangled qubits, with one qubit stored at the switch and the other at the user. Once the switch has established the required LLEs with the users involved in a request (where a request specifies a subset of users that wish to establish entanglement among them through the switch), it performs a swapping operation \cite{valls2024brief} on the stored qubits at the switch that, if successful, converts these LLEs into end-to-end entanglement between the remote parties, successfully serving the corresponding request. An illustrative example is provided in Fig.~\ref{fig:swapping}.

Previous works on quantum switches have studied their performance under a variety of network settings and with respect to various metrics. In \cite{vardoyan2019stochastic, vardoyan2020exact, nain2020analysis}, a policy determining the actions of the switch is fixed, and closed-form expressions are obtained for the switch capacity, defined as the number of end-to-end entanglements served per unit time, and for the expected number of quantum memory registers required to store qubits associated with LLEs at the switch. In \cite{dai2022capacity, vasantam2022throughput, promponas2024maximizing}, the focus is on queue stability, where they design throughput-optimal policies under different network settings. Additionally, the fidelity between entangled quantum states, a measure of similarity between quantum states which \cite{nielsen2010quantum} used to characterize the quality of entanglement, has also been adopted as a performance metric in \cite{panigrahy2023capacity, jia2024fidelity}.

This work is inspired in part by \cite{promponas2024maximizing}, which motivates our adoption of a hard memory constraint at the switch; although our setting differs in important ways. In brief, the authors of \cite{promponas2024maximizing} consider a quantum switch which serves multipartite entanglement requests under a strict memory constraint, that is, the switch must decide at the beginning of each slot how the limited memory registers should be allocated to users so that LLEs can be established. They propose a throughput-optimal policy but its worst-case complexity is prohibitive. Specifically, the authors note that, it needs to solve, in the worst case, an exponential number of NP-hard problems for every memory allocation. This naturally raises the following question, which we address in this work:\\

\emph{For a memory-constrained quantum switch, can we design implementable scheduling policies with theoretical guarantees that directly translate to quality-of-service (QoS) guarantees?} \\

Motivated by the age of information (AoI) literature \cite{yates2021age}, we adopt an age-based metric so as to evaluate the performance of a quantum switch under the proposed scheduling policies. Note that, serving a request in this setting corresponds to establishing an entanglement among the parties through the switch. In \cite{mitrolaris2025age}, we studied a job-offloading system and introduced the metric of \emph{age of job completion (AoJC)}. Inspired by that, in this work, we define a new similarly-motivated metric, the \emph{age of entanglement establishment (AoEE)}.  For each request $r$, the AoEE is defined as the time elapsed since the most recent slot in which $r$ was successfully served. 

This is a simple yet useful metric, since knowing the average AoEE, we can provide users with a quantitative estimate of how long, on average, they must wait before a given request is served again. In other words, theoretical guarantees under this metric directly translate into meaningful QoS guarantees. Moreover, even though AoEE is not a pure throughput metric, minimizing it naturally leads to shorter inter-service times, and consequently, to an increased number of end-to-end entanglements served per unit time. AoI has already been considered in the quantum communication literature, e.g., see~\cite{badia2023strategic, jabrayilova2025age}. Those works considered a strategic communication system and found game-theoretic equilibria. On the contrary, in this paper, we consider a scheduling problem where minimizing AoEE acts as a QoS objective for a quantum switch. 

In this work, we consider a time-slotted model where a switch, capable of serving multipartite entanglement requests, is connected to $N$ users. There are in total $R$ requests, with each one of them specifying a distinct subset of users that wish to establish an entanglement. The switch is assumed to operate under full-load, hence all requests are always available for scheduling. Scheduling is challenging because the switch has a hard memory constraint. In particular, at the beginning of every slot, the switch has to allocate the $M$ available quantum memory registers for storing qubits associated with the LLEs.

To obtain policies that can be employed in practical scenarios, the system model considered in this work is general in the sense that we do not make assumptions about the memory size $M$. This model, as explained in Section~\ref{sec:results}, encompasses as a special case, scenarios where there are exponentially many (in $N$) requests that need to be scheduled while the available memory is limited. Further, both LLE generation and entanglement swapping are probabilistic. At the end of each slot, any unused qubits in the memory are assumed to decohere and are thus discarded. 

The contributions of this work are as follows. We introduce the AoEE, an age-based metric for evaluating the performance of quantum switches beyond throughput, which was the main focus of most prior work. In the general setting of our system model, we develop three implementable scheduling policies and provide performance analysis for each of them. The policies are implementable in the sense that, in each slot, they can be executed with a number of computations linear in the number of requests. Importantly, the proposed policies can be deployed for any memory size and any set of requests, making them good baseline candidates for future work across different network models. Additionally, we provide numerical comparisons of the proposed policies studying the effects of memory size and set of requests.

To the best of our knowledge, this is the first line of work considering an age-based metric for the evaluation of the performance of a quantum switch. Due to space limitations, in this work, we only study three policies. Extending on this line of work, we devise index-based policies in a subsequent paper \cite{banerjee2026index}. We also omit the proofs of our results here due to space limitations, which will be provided in the journal version.

\section{System Model}
We consider a quantum switch that serves $N$ users arranged in a star topology. Let the set of requests be $\mathcal{R} = \{1, 2, \ldots, R\}$, where each request  $r \in \mathcal{R}$ uniquely specifies a distinct subset of users who intend to share entanglement. For each $r \in \mathcal{R}$, denote by $\mathcal{G}(r) \subseteq \{1,2, \ldots, N\}$ the subset of users associated with request $r$. We refer to the number of users participating in request $r$, $\left| \mathcal{G}(r)\right|$, as the cardinality of $r$. Let $\Lambda \equiv \Lambda(\mathcal{R}) = \{ \left|\mathcal{G}(r)\right|: r \in \mathcal{R}\}$ denote the set of distinct request cardinalities. For notational convenience, suppose $\Lambda = \{ \lambda_1, \lambda_2, \ldots, \lambda_k\}$ and define $\mathcal{C}(\lambda) = \{ r\in \mathcal{R}:   \left| \mathcal{G}(r)\right|=\lambda\} $, where $\lambda\in\Lambda$, as the set of requests with cardinality $\lambda$. 

The switch operates in slotted-time and has $M$ quantum registers available for storing LLEs.  We  assume $M$ is large enough to accommodate the request involving the most users, and small enough so that not all requests can be scheduled simultaneously.
Due to this memory constraint, only certain combinations of requests can be served simultaneously. At the beginning of every slot, the switch selects a subset of requests such that the total number of required registers does not exceed $M$, i.e., the sum of their cardinalities is at most $M$. We refer to such memory configurations as admissible. The set of admissible memory configurations is given by 
\begin{align}
    \mathcal{X} \!=\! \bigcup_{\ell \geq 1} \Big\{ (r_1, \ldots, r_\ell): r_i\in \mathcal{R}, \;  r_i \neq r_j, \sum_{j=1}^\ell\left|\mathcal{G}(r_j) \right| \leq M \Big\}.
\end{align}
For each slot $t$, a policy $\bm \pi$ specifies which memory configuration is selected, which we denote with $x^{\bm \pi}(t)\in\mathcal{X}$. Once an admissible memory configuration for slot $t$ has been decided, the switch and the users attempt to establish the necessary LLEs. Let $u_r^{\bm \pi}(t) \in \{0,1\}$ be equal to $1$ when $r \in x^{\bm \pi}(t)$, i.e., when request $r$ is selected for service at slot $t$, and $0$ otherwise. If $u^{\bm \pi}_r(t) = 1$, then $\left| \mathcal{G}(r) \right|$ memory registers are allocated for request $r$ so that each user $i \in \mathcal{G}(r)$ can attempt to establish an LLE with the switch for participating in request $r$. Each user $i$ successfully connects with probability $p_i \in (0,1]$, independent of the other users and attempts.

Note that a user $i$ may attempt to establish multiple LLEs with the switch if multiple scheduled requests include user $i$. To model this behavior and the associated randomness in LLE outcomes, we define the following random variables. For each request $r\in \mathcal{R}$ and user $i \in \mathcal{G}(r)$, let $b_{r,i}^{\bm \pi}(t) \in \{0,1 \}$ be equal to $1$ if, at slot $t$,  request $r$ has been selected for service under policy $\bm \pi$ (i.e., $u^{\bm \pi}_r(t)=1$) and user $i$ successfully establishes an LLE with the switch for that request; otherwise, $b_{r,i}^{\bm \pi}(t) =0$. Conditioned on $u^{\bm \pi}_r(t)=1$, $b_{r,i}^{\bm \pi}(t)$ takes the value of $1$ with probability $p_i$ and the value of $0$ with probability $1-p_i$. If  $u^{\bm \pi}_r(t)=0$, then $b_{r,i}^{\bm \pi}(t) =0$.

Let $c^{\bm \pi}_r(t) \in \{0,1\}$ indicate whether all the users in $\mathcal{G}(r)$  establish LLEs with the switch at slot $t$, i.e., $c^{\bm \pi}_r(t) = \prod_{j \in \mathcal{G}(r)} b_{r,j}^{\bm \pi}(t)$. Since LLEs are established independently, $c^{\bm \pi}_r(t)=1$ with probability $v(r) := \prod_{j \in \mathcal{G}(r)}p_j$, and only when $u_r^{\bm \pi}(t) =1$. If $c^{\bm \pi}_r(t) = 0$ and $u_r^{\bm \pi}(t) =1$, at least one user in $\mathcal{G}(r)$ fails to establish an LLE, and consequently request $r$ is not served in slot $t$. In this case, a non-empty subset of users in $\mathcal{G}(r)$ may still have successfully established LLEs, $b_{r,j}^{\bm \pi}(t) =1$ for some $j \in \mathcal{G}(r)$. Any such LLEs remain unused (i.e., they are not repurposed to serve other requests) and are discarded at the end of the slot. Any practical system that reuses such successfully established LLEs can only improve performance, thus the performance guarantees derived in this work can be interpreted as upper bounds on the achievable average age.

If $c^{\bm \pi}_r(t) = 1$, the switch attempts to generate end-to-end entanglement among the users in $\mathcal{G}(r)$. This operation is probabilistic and succeeds with a non-zero probability, which depends on the number of users involved in the request. If the cardinality of request $r$ is $\lambda$, then the success probability is $q_\lambda \in (0,1]$.

Let $d^{\bm \pi}_r(t) \in \{ 0,1\}$  denote whether request $r$ is served in slot $t$. Then, 
\begin{align}
    d^{\bm \pi}_r(t) = 
    \begin{cases}
    0,  & \text{if } u^{\bm \pi}_r(t) =0,  \\
    0, & \text{if } u^{\bm \pi}_r(t) =1 \: \text{ w.p.} \; 1-q_\lambda v(r) ,\\
    1, & \text{if } u^{\bm \pi}_r(t) =1 \: \text{ w.p.} \;  q_\lambda v(r).
    \end{cases}
\end{align}

Formally, we define the AoEE, or simply the age, for request $r$ at time $t$ as,
\begin{align}
    h^{\bm \pi}_r(t) = t - \sup\{ \tau \in \{1,2,\ldots,t-1\}:  \; d^{\bm \pi}_r(\tau) = 1 \}.
\end{align}
In particular, if $r$ has never been successfully served before slot $t$, the supremum is $0$, and thus $h_r^{\bm{\pi}}(t) = t$. The long-term time-average expected age of request $r$ under policy $\bm \pi$ is defined as,
\begin{align}
    \Delta^{\bm \pi}_r = \limsup_{T \rightarrow \infty}\frac{1}{T} \sum_{t=1}^T \mathbb{E}[h^{\bm \pi}_r(t)],
\end{align}
which we refer to as the average age of request $r$. 

In this work, our objective is to analyze and design policies that minimize the overall average age,
\begin{align}
\Delta^{\bm \pi} = \frac{1}{R}\sum_{r=1}^R \Delta^{\bm \pi}_r.
\end{align}
Thus, the optimization problem of interest is,
\begin{align}
    \inf_{\bm{\pi}\in\Pi} \ \limsup_{T \rightarrow \infty} \frac{1}{TR}\sum_{r = 1}^R \sum_{t = 1}^T \mathbb{E}\left[ h_r^{\bm \pi}(t)\right] \label{eq:n2},
\end{align}
where $\Pi$ is a set of all causal policies that satisfy the memory constraint at every slot.

\section{Main Results} \label{sec:results}
First, consider the special case in which all requests are bipartite, i.e., $\Lambda = \{2\}$. In this case, a natural policy is to fix a probability distribution over $\mathcal{R}$, and in each slot, sample requests without replacement according to this distribution until the memory is filled, i.e., sample $\lfloor \frac{M}{2} \rfloor$ requests. This yields what is typically referred to as a stationary randomized policy \cite{kadota2019minimizing, neely2010stochastic}, since the same sampling distribution is used in every
slot.

Our setting allows for the case where we may have all possible requests with cardinalities ranging from $2$ up to $N$, i.e., the cardinality of the corresponding $\mathcal{R}$ is,
\begin{align}
|\mathcal{R}| \;=\; \sum_{\lambda = 2}^N \binom{N}{\lambda},
\end{align}
which is equal to $2^N-(N+1)$ and already grows exponentially in $N$. The set of admissible memory configurations $\mathcal{X}$ then has cardinality that grows combinatorially with $|\mathcal{R}|$ and $M$. A stationary randomized policy would require specifying a probability for each $x \in \mathcal{X}$, which is infeasible in this general setting. This motivates our focus on structured, low-complexity policy families that are inspired by policies encountered in the AoI literature.

\subsection{Single-Cardinality Stationary Randomized Scheduling Policies} \label{subsec:sr}
The first family of policies we propose, termed  single-cardinality stationary randomized (SSR) scheduling policies, relies on the idea that for a fixed cardinality $\lambda \in \Lambda$ we can employ a stationary randomized policy over the corresponding requests $\mathcal{C}(\lambda)$. In each slot, the policy proceeds in two steps. First, the switch randomly selects a cardinality $\lambda \in \Lambda$ and decides to schedule only those requests whose cardinality equals $\lambda$. Then, sampling without replacement from the corresponding set $\mathcal{C}(\lambda)$, the switch schedules as many requests as possible, namely, $M_\lambda := \min \{ \left| \mathcal{C}(\lambda) \right|,  \lfloor \frac{M}{\lambda} \rfloor \}$, and proceeds to attempt servicing the selected requests. 

\begin{definition} \label{def:1}
    A single-cardinality stationary randomized scheduling policy (SSR) $\bm \mu$ is specified by a tuple $\left( \mu^{(0)}, \mu^{(\lambda_1)}, \mu^{(\lambda_2)}, \ldots, \mu^{(\lambda_k)}\right)$, where $\mu^{(0)}$ is a probability distribution supported on $\Lambda = \{\lambda_1, \lambda_2, \ldots, \lambda_k \}$, and for each $\lambda_i \in \Lambda$, $\mu^{(\lambda_i)}$ is a measure supported on $\mathcal{C}(\lambda_i)$ such that $\mu^{(\lambda_i)}(r)$, for $r \in \mathcal{C}(\lambda_i)$, represents the marginal selection probability of request $r$ and satisfies, 
    \begin{align}
        \mu^{(\lambda_i)}(r) \in [0, 1], \ r\in \mathcal{C}(\lambda_i), \quad \sum_{r \in \mathcal{C}(\lambda_i)} \mu^{(\lambda_i)}(r) = M_{\lambda_i}.
    \end{align}
    At each time slot $t$, the switch samples a cardinality value $\lambda_i$ from $\mu^{(0)}$. Conditioned on the sampled value $\lambda_i$, it considers the set of requests whose cardinality equals $\lambda_i$, $\mathcal{C}(\lambda_i)$, and selects $M_{\lambda_i}$ requests from this set without replacement using a sampling procedure whose marginal probabilities coincide with the measure $\mu^{(\lambda_i)}$. 
\end{definition}

In our theoretical analysis, we are only concerned with the marginal selection probabilities induced by the sampling procedure. For this reason, in Definition~\ref{def:1}, we do not specify the full sampling procedure but only the marginals $\mu^{(\lambda_i)}$. An efficient sampling procedure for this problem can be found in the literature, for instance, for a linear (in $|\mathcal{C}(\lambda)|$) time algorithm  \cite{gandhi2006dependent}.    

\begin{algorithm}[t]
\caption{Monotone search for finding $\gamma^{(\lambda)}$}
\label{alg:2}
\begin{algorithmic}[1]
    \STATE $\gamma^{(\lambda)} \gets \min_{r \in \mathcal{C}(\lambda)} 1/v(r)$
    \STATE $g_\lambda(\gamma^{(\lambda)}) \gets \sum_{r \in \mathcal{C}(\lambda)} \min \left\{ 1, \frac{1}{\sqrt{v(r) \gamma^{(\lambda)}}}\right\}$
    
    \WHILE{$g_\lambda(\gamma^{(\lambda)}) > M_\lambda$}
        \STATE increase $\gamma^{(\lambda)}$ by a small amount

        \STATE $g_\lambda(\gamma^{(\lambda)}) \gets \sum_{r \in \mathcal{C}(\lambda)} \min \left\{ 1, \frac{1}{\sqrt{v(r) \gamma^{(\lambda)}}}\right\}$
    \ENDWHILE
    
    \RETURN $\gamma^{(\lambda)}$
\end{algorithmic}
\end{algorithm}

We denote the collection of all SSR policies by $\Pi_{\text{\scriptsize SSR}}$. The following lemma characterizes their performance.

\begin{lemma} \label{lem:1}
    For any policy $\boldsymbol{\mu} \in \Pi_{\text{\scriptsize SSR}},$ the average age is given by 
    \begin{align}
        \Delta^{\boldsymbol{\mu}} = \frac{1}{R} \sum_{\lambda \in \Lambda} \frac{1}{\mu^{(0)}(\lambda) q_\lambda}\sum_{r \in \mathcal{C}(\lambda)} \frac{1}{\mu^{(\lambda)}(r) v(r)}. \label{eq:1}
    \end{align}
\end{lemma}

Using Lemma~\ref{lem:1}, we obtain a closed-form expression for the optimal SSR policy, stated in the next theorem.

\begin{theorem} \label{thm:1}
    Let $\bm{\mu}^\star = \left( \mu^{\star(0)}, \mu^{\star(\lambda_1)}, \mu^{\star(\lambda_2)}, \ldots, \mu^{\star(\lambda_k)}\right)$ denote the optimal SSR policy. Then, for each $\lambda \in \Lambda$ and $r \in \mathcal{C}(\lambda)$,
    \begin{align}
        \mu^{\star(\lambda)}(r) = 
        \begin{cases} 
             \min \left\{1, \frac{1}{\sqrt{\gamma^{(\lambda)} v(r)}} \right\}, & \text{if} \: \: \: M_{\lambda} < \left| \mathcal{C}( \lambda)\right| \\
            1, & \text{if} \: \: \: M_{\lambda} = \left| \mathcal{C}( \lambda)\right| ,
        \end{cases} \label{eq:6s}
    \end{align}
    where $\gamma^{(\lambda)}$ is chosen such that $\sum_{r \in \mathcal{C}(\lambda)} \min \left\{1, \frac{1}{{\sqrt{\gamma^{(\lambda)} v(r)}}} \right\} = M_\lambda$, and can be computed by Algorithm~\ref{alg:2}. Moreover, the optimal cardinality-selection probabilities are given by,
    \begin{align}
        \mu^{\star(0)}(\lambda) = \frac{\sqrt{f_\lambda^\star/q_\lambda }}{ \sum_{\lambda' \in \Lambda} \sqrt{ f_{\lambda'}^\star/q_{\lambda'}}}, \label{eq:7s}
    \end{align}
    where, $f_\lambda^\star := \sum_{r \in \mathcal{C}(\lambda)} \frac{1}{\mu^{\star(\lambda)}(r) v(r)}$.
\end{theorem}

\subsection{Single-Cardinality Max-Weight Scheduling Policy} \label{subsec:smw}
It is common in the AoI literature to first develop a stationary randomized policy and then use Lyapunov optimization to construct age-based max-weight policies \cite{kadota2019minimizing}. In what follows, we adopt a similar approach tailored to our setting.

As in the SSR family, the switch first selects a cardinality $\lambda$, according to some probability distribution, and restricts scheduling decisions to requests in $\mathcal{C}(\lambda)$. However, instead of  determining the specific requests to be allocated in the memory with a stationary randomized policy, the switch selects them according to a max-weight rule.

\begin{definition}
    Under the single-cardinality max-weight (SMW) policy $\bm\psi$, at each slot $t$, the switch samples a cardinality value $\lambda$ from $\mu^{\star(0)}$ specified in \eqref{eq:7s}. Conditioned on the sampled value $\lambda$, it considers the set of requests whose cardinality equals $\lambda$, $\mathcal{C}(\lambda)$, and selects the $M_{\lambda}$ requests in this set with the largest weights $h^{\bm\psi}_r(t)/\mu^{\star(\lambda)}(r)$, with $\mu^{\star(\lambda)}(r)$ specified in \eqref{eq:6s}. Then, the memory configuration at slot $t$, conditioned on scheduling requests of cardinality $\lambda$, is given by,
    \begin{align}
        x^{\bm\psi}(t) = \arg \max_{ \substack{C \subseteq \mathcal{C}(\lambda) \\ \lvert C \rvert = M_{\lambda}}} \sum_{r \in C} \frac{ h_r^{\bm \psi}(t)}{\mu^{\star(\lambda)}(r)},
    \end{align}
    with ties broken arbitrarily.
\end{definition}

Both the SMW policy $\bm \psi$ and the optimal SSR policy $\bm \mu^{\star}$ use the same probability distribution to determine the cardinality of the requests to be scheduled.   The use of the max-weight rule in SMW, rather than randomizing without accounting for the current state information, leads to improved performance as formally established in the next theorem.

\begin{theorem} \label{thm:2}
    The SMW policy $\bm\psi$ performs no worse than the optimal SSR policy $\bm\mu^\star$,
    \begin{align}
        \Delta^{\bm\psi} \leq \Delta^{\bm\mu^\star}.
    \end{align}
\end{theorem}

\subsection{Multi-Cardinality Max-Age Scheduling Policies}\label{subsec:mma}

Both policies developed so far rely on the same first step: randomly selecting a single cardinality $\lambda \in \Lambda$ and scheduling only requests of that cardinality in a slot. It is natural to think that performance could be improved by allowing requests with different cardinalities to be scheduled simultaneously. However, as already discussed above, considering all admissible memory configurations in this general setting is not feasible. 

To obtain a concrete tractable policy class, we consider maximal feasible subsets of cardinalities, $\Lambda^{(1)}, \Lambda^{(2)}, \ldots, \Lambda^{(n)} \subseteq \Lambda$. In particular, for every $i \in [n]:= \{1,2,\ldots,n\}$, $\sum_{\lambda \in \Lambda^{(i)}} \lambda \leq M$ and $\sum_{\lambda \in \Lambda^{(i)}} \lambda + \lambda' > M$, $\lambda'\in{\Lambda}\setminus\Lambda^{(i)}$. Here $n$ and the specific cardinality subsets are  hyper-parameters of the policy.
Note that, the total number of maximal feasible subsets depends on $\Lambda$ and $M$.  

Choosing one of these subsets based on a stationary randomized rule is the first step of the policies introduced in this subsection. Each $\Lambda^{(i)}$ specifies only the cardinalities of the requests to be allocated in a slot and does not identify specific requests. As a second step, the policy adopts the following decision rule. Conditioned on having chosen subset $\Lambda^{(i)}$, the specific requests are determined by selecting, for each cardinality $\lambda \in \Lambda^{(i)}$, the request in $\mathcal{C}(\lambda)$ with the maximum AoEE. We formally define this class of policies next.
 
\begin{definition} \label{def:2}
    For a given collection of maximal feasible cardinality subsets $\{\Lambda^{(i)}\}_{i\in [n]}$, a \emph{multi-cardinality max-age} (MMA) policy $\bm{\phi}$ is specified by a probability distribution $\phi$ supported on the index set $[n]$.
    At each time slot $t$, the switch samples an index $j$ from $\phi$. Conditioned on the sampled value $j$, it considers the cardinality subset $\Lambda^{(j)}$. For each cardinality $\lambda \in \Lambda^{(j)}$, the switch selects the request of cardinality $\lambda$  with the largest AoEE, 
    \begin{align}
        \tilde{r}_\lambda(t) := \arg \max_{r' \in \mathcal{C}(\lambda)} h^{\bm{\phi}}_{r'}(t),
    \end{align}
    and configures the memory for slot $t$ with all selected requests,
    \begin{align}
        x^{\bm{\phi}}(t) = \{ \tilde{r}_\lambda(t): \: \lambda \in \Lambda^{(j)}\}.
    \end{align}
\end{definition}

For given cardinality subsets $\{ \Lambda^{(i)}\}_{i \in [n]}$, we denote the collection of all MMA policies by $\Pi_{\text{\scriptsize MMA}}(\{ \Lambda^{(i)}\}_{i \in [n]})$. The following lemma characterizes their performance.

\begin{lemma} \label{lem:2} 
    For any policy $\bm{\phi} \in \Pi_{\text{\scriptsize MMA}}(\{ \Lambda^{(i)}\}_{i \in [n]})$, the average age of a request $r \in \mathcal{C}(\lambda)$ is given by,
    \begin{align}
        \Delta_{r}^{\bm \phi} = \frac{1}{2\theta_{\lambda} q_{\lambda}} \left(\frac{1}{\beta_{ \lambda}}\sum_{r' \in \mathcal{C}(\lambda)} \frac{1}{v(r')^2} + \beta_{\lambda} \right),   \label{eq:13s}
    \end{align}
    where $\theta_\lambda := \sum_{i: \lambda \in \Lambda^{(i)}}\phi(i)$ and $\beta_\lambda := \sum_{r' \in \mathcal{C}(\lambda)} \frac{1}{v(r')}$.
\end{lemma}

From Lemma~\ref{lem:2} we see that requests of the same cardinality have the same average AoEE under an MMA policy. This is due to the round-robin scheduling in each $\mathcal{C}(\lambda)$ resulting from the max-age selection rule. This fact can be interpreted as an advantage of this policy since it ensures that requests of the same cardinality are treated equally, that is, they have the same average age. However, a downside of the round-robin scheduling per $\mathcal{C}(\lambda)$ is that the performance of the policy can be compromised by requests with low probability of successful service. Once a struggling request has max-age in $\mathcal{C}(\lambda)$, it keeps getting scheduled whenever $\lambda$ is chosen until it is serviced successfully since its age remains the largest.

Lemma~\ref{lem:2} implies the following convex optimization problem for the optimal sampling distribution $\phi^\star$ over cardinality subsets.

\begin{prop}
    The optimal policy $\bm \phi^\star \in \Pi_{\text{\scriptsize MMA}}(\{ \Lambda^{(i)}\}_{i \in [n]})$ employs the sampling distribution $\phi^\star$ given by,
    \begin{align}
       \phi^\star \ \in \   &\arg\min_{\phi} \ \sum_{\lambda \in \Lambda}\frac{1}{\sum_{i: \lambda \in \Lambda^{(i)}} \phi(i)} w_\lambda \nonumber \\
        & \ \text{s.t.} \qquad \ \phi(j) \geq 0, \ \sum_{j \in [n]} \phi(j) = 1,
    \end{align}
    where $w_\lambda := \frac{|\mathcal{C}(\lambda)|}{2q_\lambda} \left(\frac{1}{\beta_\lambda} \sum_{r \in \mathcal{C}(\lambda)}\frac{1}{v(r)^2} + \beta_\lambda \right)$.
\end{prop}

\section{Numerical Results}
\begin{figure}[t]
    \centering
    \includegraphics[width=0.95\linewidth]{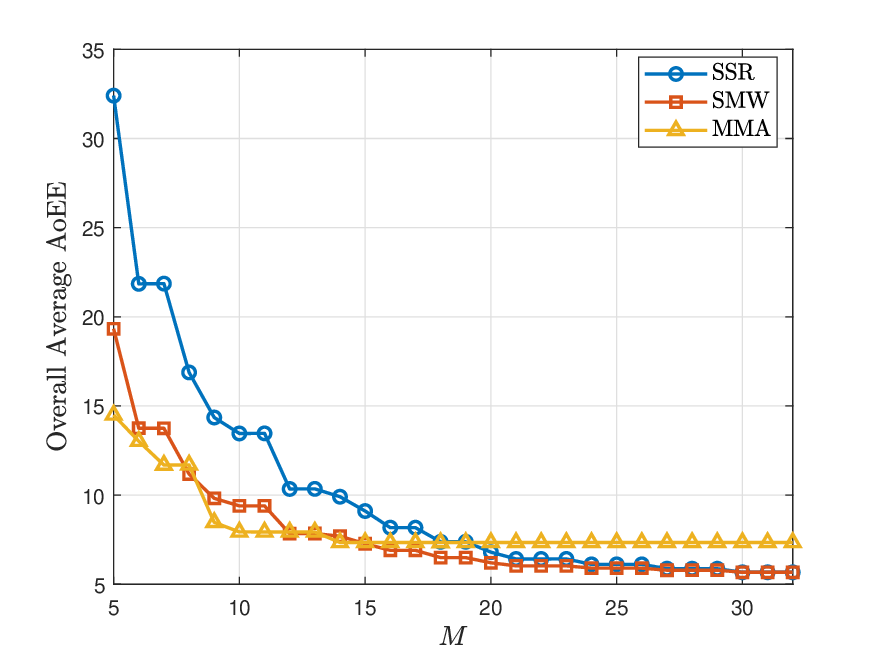}
    \caption{Average AoEE achieved by the proposed policies as a function of the memory size $M$ in a network with $N=5$ users and all possible requests, $|\mathcal{R}|=2^N-(N+1)=26$.}
    \label{fig:plot1}
\end{figure}

In this section, we numerically compare the performances of the three proposed policies in terms of the average AoEE. We first study the effect of the memory size $M$. We consider a network with $N=5$ users and all possible requests (i.e., all user subsets of size at least two), thus the total number of requests is $|\mathcal{R}| = 2^N-(N+1)=26$. Users establish LLEs with probabilities $p_1=0.85$,  $p_2 = 0.9$, $p_3 = 0.93$, $p_4 = 0.87$ and $p_5 = 0.95$, whereas the swapping probabilities are selected as $q_2 = 0.92$, $q_3 = 0.87$, $q_4  =0.83$ and $q_5 = 0.8$. In  Fig.~\ref{fig:plot1}, we evaluate the average AoEE achieved by each policy as the memory size $M$ increases from $M=5$ to $M=32$. For the MMA policy, we consider all possible maximal feasible cardinality subsets for each value of $M$. For small values of $M$, MMA outperforms the SSR and SMW. This is because when $M$ is small enough, e.g., $M=5$, SSR and SMW can schedule either two bipartite requests or a single request of greater cardinality, leaving memory registers unused. In contrast, MMA makes better use of the memory by scheduling requests of cardinality two and three simultaneously.

As $M$ increases, the performances of all three policies improve. Interestingly, MMA saturates for $M\geq 14$. Note that, when $M \geq 14$, the only maximal feasible cardinality subset is $\Lambda$ itself, which uses $\sum_{\lambda \in \Lambda}\lambda =14$ memory registers. As a result, the policy only uses $14$ memory registers even when $M \geq 14$, thus explaining the constant performance for $M \geq 14$. This highlights a limitation of the MMA policy, namely that, in systems with a large number of quantum memory registers, it does not make use of all available resources, hence, hindering the overall performance. 
     
As expected from Theorem~\ref{thm:2}, the SMW policy performs at least as good as the SSR policy, with the performance gap being larger for small $M$. With the number of memory registers increasing, the performance of the SSR gets closer to that of SMW. As $M$ increases, both SSR and SMW schedule more requests, leading to a greater overlap in their selections since both policies employ the same probability distribution for sampling cardinalities. In fact, the performance of the SSR exactly matches that of SMW when $M \geq \max_{\lambda \in \Lambda} \lambda |\mathcal{C}(\lambda)|$, in which case the memory is large enough to accommodate all requests of any sampled cardinality. In Fig.~\ref{fig:plot1}, this happens when $M=30$.  

Additionally, the performance curves in Fig.~\ref{fig:plot1} exhibit a stepwise behavior. For instance, increasing $M$ from five to six significantly affects the performance of both SSR and SMW, but increasing $M$ from six to seven yields no improvement. This is because, the number of same-cardinality requests that can be scheduled per slot is $M_\lambda = \min\{|\mathcal{C}(\lambda)|, 
\lfloor M/\lambda\rfloor\}$, which changes only when $\lfloor M/\lambda\rfloor$ increases. With quantum memories being a scarce resource, identifying such impactful memory increments is important. 

\begin{figure}[t]
    \centering
    \includegraphics[width=0.95\linewidth]{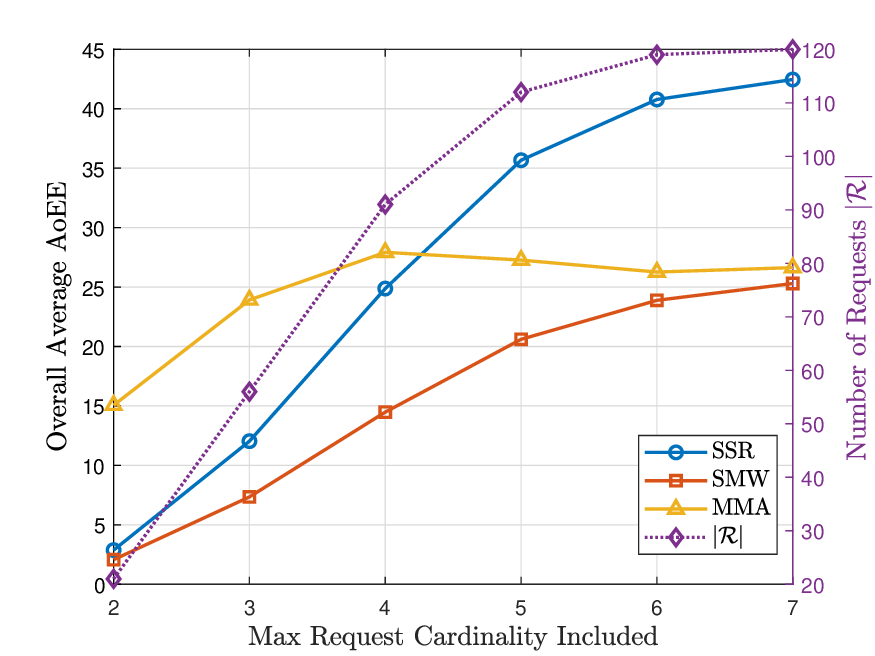}
    \caption{Average AoEE achieved by the proposed policies as the set of requests expands in a network with $N=7$ users and  $M=20$. Starting from all bipartite requests, we progressively include all requests of larger cardinalities until all possible requests are present.}
    \label{fig:plot2}
\end{figure}

Next, we study how expanding the request set affects the performance for a fixed memory size. We consider a network with $N=7$ users and $M=20$. Users establish LLEs with probabilities $p_1= 0.85$,  $p_2=0.9$, $p_3=0.93$, $p_4=0.87$, $p_5 = 0.95$, $p_6=0.83$, and $p_7=0.92$, whereas swapping probabilities are selected as $q_2 = 0.92$, $q_3 = 0.87$, $q_4  =0.83$, $q_5 = 0.8$, $q_6=0.78$ and $q_7=0.75$. In  Fig.~\ref{fig:plot2}, we evaluate the average AoEE achieved by each policy as the set of requests expands. We begin with all bipartite requests and progressively include all requests of larger cardinalities until the full request set for seven users is reached. For instance, when the maximum request cardinality included is three, the request set consists of all possible bipartite and tripartite requests. The dotted curve (right axis) shows the number of requests $|\mathcal{R}|$. For the MMA policy, we consider all possible maximal cardinality subsets in each case.

We observe that as the set of requests grows, the performances of all three policies degrade, which is expected since more requests need to be scheduled while having the same memory budget. When only bipartite requests are present, the MMA policy can schedule at most one request per slot, leading to significantly worse performance compared to SSR and SMW, which schedule 10 out of the 21 bipartite requests in each slot. MMA becomes more competitive as requests of different cardinalities are introduced, but it still does not manage to outperform MW. Moreover, the performance gap between SMW and SSR becomes larger as the set of requests expands.  

From the discussion so far, it follows that it is possible to partition the set of requests into groups and allocate dedicated portions of memory to each group. Our policies can then be applied independently within each module, improving scalability and increasing memory utilization. Designing the partitioning and memory allocation, and selecting which policy to run in each module, is an interesting direction for future work.

\bibliographystyle{unsrt}
\bibliography{references}

\end{document}